\documentclass[twocolumn]{aastex631}

\newcommand{\gobs}{g_{\rm obs}}	
\newcommand{\gbar}{g_{\rm bar}}	
\newcommand{\rhobar}{\rho_{\rm bar}}	
\newcommand{\gpls}{g_\dag}	
\newcommand{\sigmabar}{\Sigma_{\rm bar}}
\newcommand{\sigmabardisk}{\Sigma_{\rm disk}}
\usepackage{color}

\received{\today} 
\shorttitle{Cores of dwarf galaxies in MOND}
\shortauthors{S\'anchez Almeida}
\graphicspath{{./}}

\begin{document}

\title{
  Dwarf galaxies with central cores in modified Newtonian dynamics (MOND) gravity}

\correspondingauthor{JSA}
\email{jos@iac.es}

\author[0000-0003-1123-6003]{J. S\'anchez Almeida} \affil{Instituto de Astrof\'\i sica de Canarias, La Laguna, Tenerife, E-38200, Spain} \affil{Departamento de Astrof\'\i sica, Universidad de La Laguna, Spain}



\begin{abstract}
Some dwarf galaxies are within the Mondian regime  at all radii, i.e., the gravitational acceleration provided by the observed baryons is always below the threshold of $g_\dag\simeq 1.2\times 10^{-10}\,{\rm m\,s^{-2}}$. These dwarf galaxies often show cores, in the sense that assuming Newton's gravity to explain their rotation curves, the total density profile $\rho(r)$ presents a central plateau or {\em core} ($d\log \rho/d\log r\rightarrow 0$ when $r\rightarrow 0$). Here we show that under MOND gravity, the existence of this core implies a baryon content whose density $\rhobar$ must decrease toward the center of the gravitational potential ($\rhobar\rightarrow 0$ when $r\rightarrow 0$). Such drop of baryons toward the central region is neither observed nor appears in numerical simulations of galaxy formation following MOND gravity. We analyze the problem posed for MOND as well as possible workarounds.

\end{abstract}




\section{Introduction}\label{sec:intro}

The existence of dark matter (DM) and its nature are probably two of the main unsolved problems in physics \citep[e.g.,][]{1987ARA&A..25..425T,2000RPPh...63..793B,2011ARA&A..49..155P,2015PDU.....7...16B,2021arXiv210602672P}. Among the solutions, \citet{1983ApJ...270..365M} circumvented the need for DM by modifying Newton's gravitational law at very low gravities, a workaround that is able to explain some of the issues that DM explains, in  particular, the rotation curves (RC) in the outer parts of regular galaxies.  \citeauthor{1983ApJ...270..365M}'s  phenomenological theory is known as MOND (MOdified Newtonian Dynamics) and has received much attention over the years \citep[][]{2001AcPPB..32.3613M,2002ARA&A..40..263S,2012LRR....15...10F,2021PhRvL.127p1302S}. It is also known to face some difficulties, though
\citep[e.g.,][]{2013MNRAS.436..202A,2014MNRAS.439.2132R,2021ApJ...914L..37S}. 
%
In this context where sensible explanations for the DM-related phenomena are eagerly sought, we have come across a theoretical result that may help us to constrain the validity of the MOND hypothesis. It is reported here.

In essence, MOND gravity establishes that Newton's law holds when the gravitational acceleration produced by baryons alone, $\gbar$, is larger than the threshold  $\gpls \simeq 1.2\times 10^{-10}\,{\rm m\,s^{-2}}$ that defines  the Mondian  regime (details and references in Sect.~\ref{sec:maineqs}). Otherwise, the gravitational acceleration scales as $\sqrt{\gpls\gbar}$, which is larger than $\gbar$ alone.  In this paper, we adopt the parameterization of MOND known as Radial Acceleration Relation  \citep[RAR;][]{2016PhRvL.117t1101M,2017ApJ...836..152L}, although our conclusions do not depend on this assumption, as we discuss in various passages of the manuscript.

Even in their centers, many dwarf galaxies with well measured RCs are in the Mondian regime. With typical central densities $\rho_{\rm co}$ of the order of $5\times 10^7\,{\rm M_\odot\,kpc^{-3}}$ and core radii $r_{\rm co}$ around 0.5\,kpc  \citep[e.g.,][Table~2]{2015AJ....149..180O}, their  gravitational acceleration is of the order of 
\begin{equation}
\frac{4\pi G}{3}\,\rho_{\rm co}\,r_{\rm co}\simeq 10^{-11}\,{\rm m\,s^{-2}},
\end{equation}
with $G$ the gravitational constant. This acceleration is clearly smaller than $\gpls$, showing that these dwarf galaxies are good testbeds for MOND gravity \citep[][]{2002ARA&A..40..263S,2019A&A...623A..36M,2022NatAs...6...35L}.

When the RC of dwarf galaxies are interpreted in terms of DM, the resulting mass density profile often shows a central plateau or {\rm core}. This originates the so-called {\em core-cusp problem} of the cold DM paradigm \citep[e.g.,][]{2017Galax...5...17D,2017ARA&A..55..343B}, since this observed core contrasts with the {\em cusp} expected in the mass density profile resulting from cold DM \citep[the NFW profile, after][]{1997ApJ...490..493N}. We will show that for this core to be consistent with RAR (and so with MOND), the baryon density should {\em decrease} toward the center of the galaxies. With all due caution, this central drop of baryons has not been observed (or, at least, it is exceptional), which questions that RAR holds in the cores of dwarf galaxies. Exceptions to RAR are to be expected within  the cold DM paradigm since RAR is view only as an emergent relation resulting from the complex interplay between baryons and DM \citep[e.g.,][]{2017MNRAS.471.1841N,2017PhRvL.118p1103L}. For MOND, however, RAR is a law of nature to be followed in every occasion, also in the cores of galaxies.
 
The paper is organized as follows:
Sect.~\ref{sec:maineqs} shows how for the baryon density profiles to be simultaneously consistent with RAR and with central cores,  their density has to decrease toward the center of the gravitational well. An approximation for 2-dimensional baryon distributions is also worked out in Sect.~\ref{sec:diskbar}.
The surface density of profiles accounting for central cores plus RAR is computed in Sect.~\ref{sec:rho2sigma}.
Section~\ref{sec:observations} discusses existing observations, which discard a significant lack of baryons in galaxy centers. In particular,  we show how the central parts of dwarfs with well-measured RCs often defy RAR. Numerical simulations of galaxies based on MOND do not show a drop of baryons in their centers, as discussed in Sect.~\ref{sec:conclusions}.   Possible workarounds for MOND to account for the difficulty posed in the paper are analyzed in Sect.~\ref{sec:conclusions}.

\section{Main equations}\label{sec:maineqs}

Empirically, the relation between the gravity provided by the observed baryons, $\gbar$, and the effective gravity explaining the dynamics, $\gobs$, is approximately given by the RAR  \citep[e.g.,][]{2016PhRvL.117t1101M,2017ApJ...836..152L,2018A&A...615A...3L},
\begin{equation}
  \gobs = \frac{\gbar}{1-\exp({-\sqrt{\gbar/\gpls}})},
  \label{eq:rar}
\end{equation}
so that $\gobs\simeq \gbar$ for $\gbar \gg\gpls$ and $\gobs\simeq\sqrt{\gbar\,\gpls}$ when $\gbar \ll\gpls$ (the latter usually called Mondian regime). This relation is taken in the paper as an operative parameterization of MOND gravity. Other parameterizations  are discussed at the end of the section. 

Within the DM paradigm, the difference between  $\gobs$ and $\gbar$ is due to the presence of a DM halo so that the total gravitational mass density, $\rho$, has contributions from baryons, $\rhobar$, and from DM, $\rho_{\rm DM}$, so that $\rho=\rhobar+\rho_{\rm DM}$. Assuming spherical symmetry,
\begin{equation}
  \gbar(r) = 4\pi G\frac{1}{r^2}\,\int_0^rx^2\rhobar(x)\,dx,
  \label{eq:gbar}
\end{equation}
and 
\begin{equation}
  \gobs(r) = 4\pi G\frac{1}{r^2}\,\int_0^rx^2\rho(x)\,dx,
  \label{eq:gobs}
\end{equation}
with $r$ the distance to the center of the gravitational potential. Equations (\ref{eq:gbar}) and (\ref{eq:gobs}) can be combined\footnote{Compute the derivative of $\log\gbar$ with $\log r$ from Eq.~(\ref{eq:gbar}). Repeat the exercise with $\log\gobs$ using  Eq.~(\ref{eq:gobs}). Then, keeping in mind that $(d\log\gobs/d\log r)/(d\log\gbar/d\log r) = d\log\gobs/d\log \gbar$, the ratio of the two derivatives renders Eq.~(\ref{eq:master}).} to get the total density $\rho(r)$ satisfying RAR for a given baryon density $\rhobar(r)$, explicitly,
\begin{equation}
  \rho(r) = \frac{\gobs(r)}{\gbar(r)}\Big[f^\prime(r)\,\rhobar(r)+\frac{\gbar(r)}{2\pi\, G\, r}\Big(1-f^\prime(r)\Big)\Big],
  \label{eq:master}
\end{equation}
with
\begin{equation}
  f^\prime = \frac{d\log\gobs}{d\log\gbar}.
  \label{eq:master2}
\end{equation}
Equation~(\ref{eq:rar}) allows $f^\prime$ to be evaluated, and it turns out to be,
\begin{equation}
  f^\prime = 1-\frac{\sqrt{\gbar/\gpls}}{2}\, \frac{\exp({-\sqrt{\gbar/\gpls}})}{1-\exp({-\sqrt{\gbar/\gpls}})}.
  \label{eq:fprime}
\end{equation}

\begin{figure*}
  \centering
\includegraphics[scale=0.8]{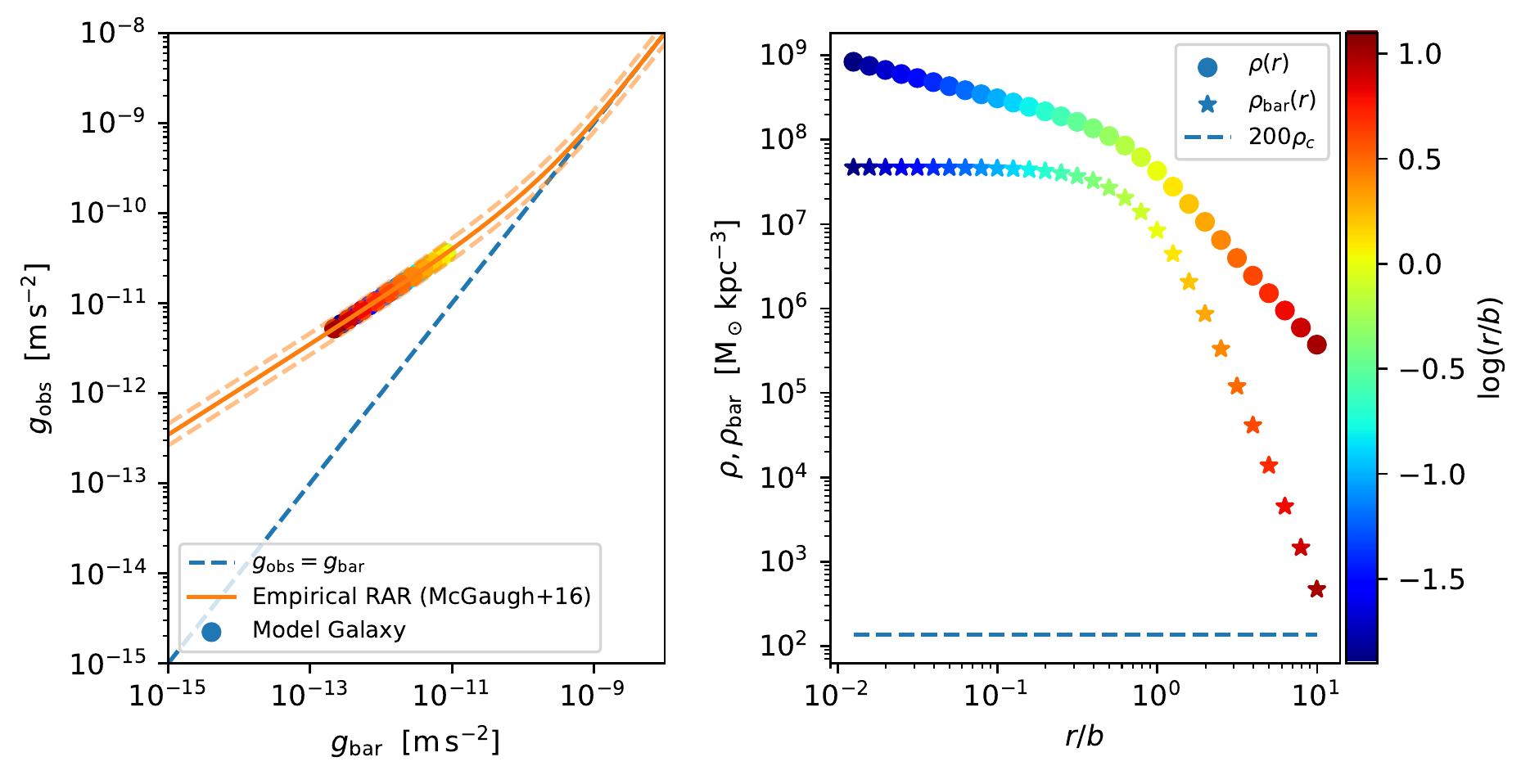}
\caption{Toy model for density profiles following RAR and having no drop in the central baryon content. Left panel: empirical relation between the radial acceleration inferred from rotation curves, $\gobs$, and the radial acceleration produced by the observed baryons, $\gbar$. This is the so-called RAR, which is shown as worked out by \citet[][the solid orange line, with the dashed lines showing the scatter]{2016PhRvL.117t1101M}. The symbols represent a model galaxy whose radial density profiles are shown in the right panel. The slanted blue dashed line shows the $x=y$ relation, and is included for reference. Right panel: baryon density (star symbols) and the total density (bullet symbols) of a model galaxy that follows RAR.
  The model galaxy has baryon mass $M_{\rm bar}=10^8\,{\rm M}_\odot$, central surface density $\Sigma_{\rm bar}(0)=50\,{\rm M}_\odot\,{\rm pc^{-2}}$, which together render a core scale length $b\simeq 0.8\,{\rm kpc}$.  
  The symbols in both the left and the right panels are color coded according to the radial distance, as given by the vertical color bar. The dashed line in the right panel indicates 200 times the critical density $\rho_c$ which usually defines the outer edge of the DM halo \citep[e.g.,][]{1997ApJ...490..493N}.}
\label{fig:rar3}
\end{figure*}

Together with  Eqs.~(\ref{eq:rar}), (\ref{eq:gbar}), and (\ref{eq:fprime}), Eq.~(\ref{eq:master}) provides the pair $\rho(r)$ -- $\rhobar(r)$ reproducing RAR. To illustrate the relation, Fig.~\ref{fig:rar3} shows $\rhobar$ given by a polytrope of index $m=5$\footnote{Also known as Plummer profile or Shutter profile: see, e.g. \citet{2008gady.book.....B}.}, which is known to reproduce the stellar mass density profiles observed in low-mass galaxies \citep[e.g.,][]{2021ApJ...921..125S}. It was chosen because this baryon density has a central plateau
($d\log\rhobar/d\log r\simeq 0$ when $r\rightarrow 0$). According to Fig.~\ref{fig:rar3}, the existence of this core in $\rhobar$ is not reflected in $\rho$, which keeps growing toward the center ($d\log\rho/d\log r\simeq -0.5$). 
\begin{figure*}
 \centering
\includegraphics[scale=0.8]{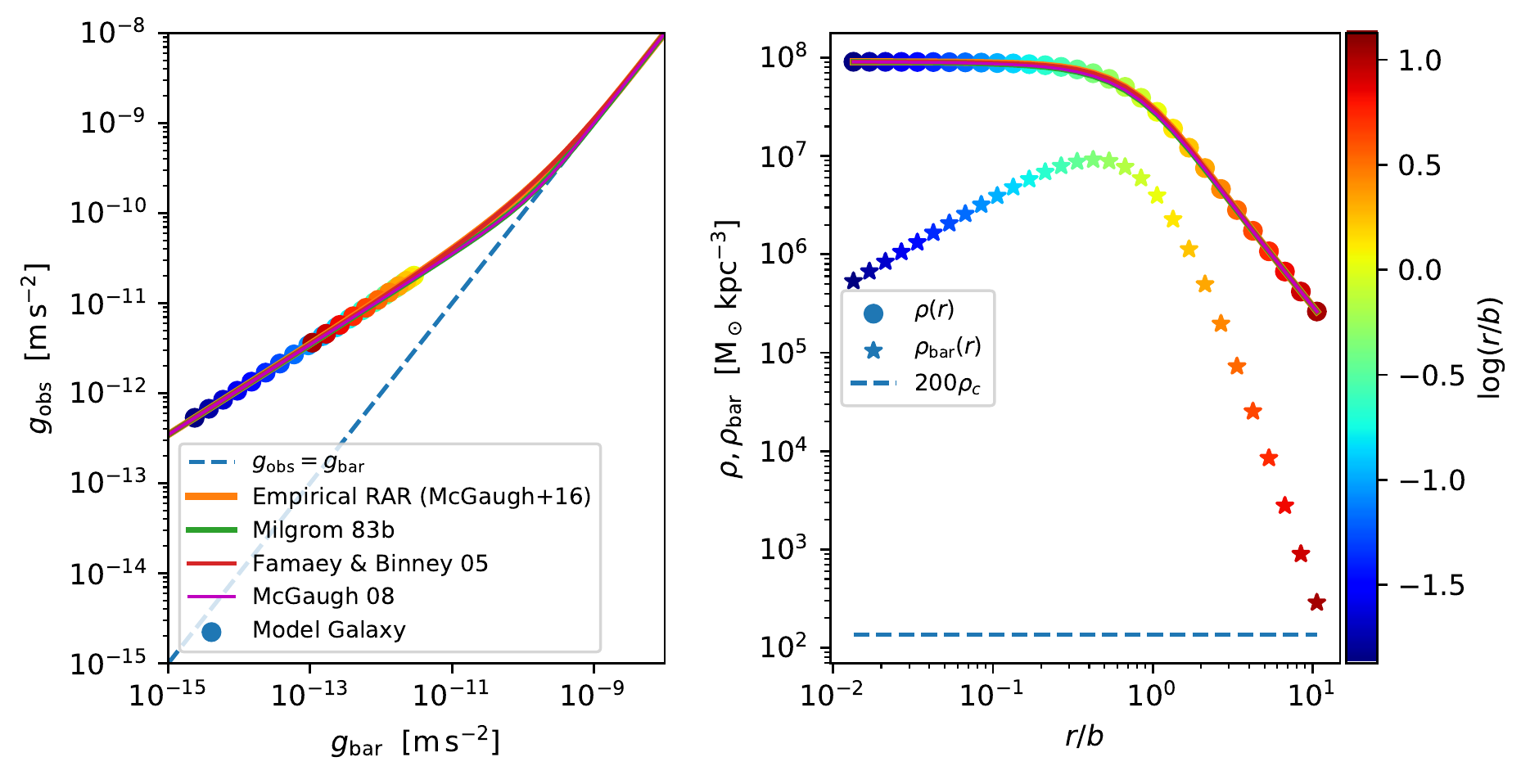}
\caption{Similar to Fig.~\ref{fig:rar3}, except that the baryon density has been chosen to vary as $\rhobar\propto r$ in the core so that the total density fulfilling the RAR (left panel) has a central plateau or {\em core} (right panel). 
  The model galaxy has stellar mass $M_{\rm bar}\simeq 7\times 10^7\,{\rm M}_\odot$ and a core scale length of  $b\simeq 0.8\,{\rm kpc}$, which render a central surface density $\Sigma_{\rm bar}(0)\simeq 17\,{\rm M}_\odot\,{\rm pc^{-2}}$. Unlike Fig.~\ref{fig:rar3},  here we show several parameterizations of RAR (left panel), which have been included to illustrate that the central core in $\rho(r)$ appears independently of the actual parameterization (see the overlapping solid lines in the right panel, with the color code given in the left panel).
}
\label{fig:rar4}
\end{figure*}

The different asymptotic behavior of $\rhobar$ and $\rho$ will hold for most baryon mass profiles provided the Mondian condition is met ($\gbar\ll\gpls$) because, if $\rhobar(r)\propto r^{-\gamma}$ when $r\rightarrow 0$, i.e., if
\begin{equation}
  \lim_{r \to 0}\, \frac{d\log\rhobar}{d\log r} = -\gamma,
\label{eq:limbar}
 \end{equation}
then Eq.~(\ref{eq:rar}) yields, 
\begin{equation}
  \lim_{r \to 0}\, \frac{d\log\rho}{d\log r} = -\frac{1+\gamma}{2}.
  \label{eq:lim}
\end{equation}
 Equation~(\ref{eq:lim}) follows from  Eq.~(\ref{eq:limbar}) through Eqs.~(\ref{eq:gbar}) and (\ref{eq:master}) for a self-gravitating system in the Mondian regime.
 Thus, for a galaxy  to have a core inferred from the rotation curve, i.e., for 
 \begin{equation}
   \lim_{r \to 0}\, \frac{d\log\rho}{d\log r} = 0,
   \label{eq:coredef}
\end{equation}
then $\gamma=-1$. As a result, the baryon density must grow linearly with radius in the center of the galaxy,
\begin{equation}
  \rhobar(r)\simeq \rhobar(r_0)\,\frac{r}{r_0},
  \label{eq:power}
\end{equation}
with $r_0$ a characteristic radius.
Figure~\ref{fig:rar4} illustrates the result. We  prescribe $\rhobar$ as a {\em hollow polytrope}, i.e.,
\begin{equation}
  \rhobar(r) = p(r/b,m) \,[1-\exp(-r/b)],
  \label{eq:hollowpoly}
\end{equation}
where $p(r/b,m)$ is a polytrope of index $m$, known to have a central core \citep[e.g.,][]{2022Univ....8..214S}. Thus, it is clear from Eq.~(\ref{eq:hollowpoly}) that $\rhobar\propto r$ when $r \ll b$, as required to reproduce a central core according to RAR ($\gamma = -1$; Eq.~[\ref{eq:lim}]). This is illustrated  with the $m=5$ hollow polytrope in Fig.~\ref{fig:rar4}, but it holds for all the other indexes as well.

Note that the main result given above, namely,  the fact that $\rhobar$ has to be proportional to $r$ for $\rho$ to have a core,  is independent of the parameterization used to represent MOND. We employ RAR (Eq.~[\ref{eq:rar}]) but several other parameterizations are available in the literature \citep[e.g.,][]{2008ApJ...678..131M,2008ApJ...683..137M}. It will hold for any of them since the result stands within the Mondian regime, which is the same for all. To further support this argument, the density profiles resulting from several parameterizations are shown in Fig.~\ref{fig:rar4}. Specifically, in addition to Eq.~(\ref{eq:rar}), Fig.~\ref{fig:rar4} includes, 
\begin{equation}
  \frac{\gobs}{\gbar}=\sqrt{\frac{1}{2}+\frac{1}{2}\sqrt{1+4/(\gbar/\gpls)^2}},
  \label{eq:appb1}
\end{equation}
from \citet{1983ApJ...270..371M},
\begin{equation}
  \frac{\gobs}{\gbar}=\frac{1}{2}+\frac{1}{2}\sqrt{1+4/(\gbar/\gpls)},
  \label{eq:appb2}
\end{equation}
from \citet{2005MNRAS.363..603F}, and
\begin{equation}
  \frac{\gobs}{\gbar}=\frac{1}{\sqrt{1-\exp(-\gbar/\gpls)}},
  \label{eq:appb3}
\end{equation}
from \citet{2008ApJ...683..137M}. The relation between $\rhobar$  and $\rho$ in Eq.~(\ref{eq:master}) remains valid in this case with the actual parameterization used for MOND entering through the derivative $f^\prime$ (Eq.~[\ref{eq:master2}]). The examples shown in Fig.~\ref{fig:rar4} employ a $f^\prime$  computed numerically from the above expressions.

%
%
\subsection{Approximation for baryons distributed in thin disks}\label{sec:diskbar}
If the baryons are in a disk (i.e., the baryon system has axi-symmetry rather than the spherical symmetry assumed in Sect.~\ref{sec:maineqs}), then Eq.~(\ref{eq:gbar}) no longer holds. The mass distribution at radii $> r$ also affects the gravitational potential at $r$ and so has to be considered to work out $\gbar$. However, considering the mass interior to $r$ still represents a sensible first order approximation \citep[][Sect.~2.6.1]{2008gady.book.....B}, which we take as an ansatz  to keep the discussion in the realm of analytical expressions. In this case,
\begin{equation}
\gbar(R) = 2\pi G\frac{1}{R^2}\,\int_0^Rx\,\sigmabardisk(x)\,dx,
  \label{eq:gbar_disk}
\end{equation}
with $\sigmabardisk(R)$ the baryon mass surface density of the disk at the radius $R$. Thus, the expression is formally identical to Eq.~(\ref{eq:gbar}) replacing
$ r$ with $R$ and 
\begin{equation}
  \rhobar(r) {\rm ~~~with~~~}  \frac{\sigmabardisk(R)}{2R}.
  \label{eq:transformation}
\end{equation}
We are interested in the central region of the galaxy. Assuming a power law approximation,
\begin{equation}
  \sigmabardisk(R) \simeq \sigmabardisk(r_0)\,(R/r_0)^{-\beta},
  \label{eq:sigmadisk}
\end{equation}
Eqs.~(\ref{eq:power}) and (\ref{eq:transformation}) imply $\gamma \rightarrow \beta+1$ so that Eq.~(\ref{eq:lim}) yields, 
\begin{equation}
  \lim_{r \to 0}~ \frac{d\log\rho}{d\log r} = -\frac{2+\beta}{2}.
  \label{eq:lim2d}
 \end{equation}
 This equation shows that for the total density to have a core (i.e., $d\log\rho/d\log r\to 0$ when $r\to 0$), the baryon
 surface density should have a central drop even more pronounced than when the mass is distributed in a 3D  volume  (i.e., rather than -1, the slope turns out to be $\beta = -2$, which corresponds to $\sigmabardisk\propto R^2$). 

The haloes resulting from DM numerical simulations are not completely spherically symmetric. However, the lower the redshift and halo mass the smaller the deviation from spherical symmetry. For example,  low-mass haloes with mass $< 10^{13}\,{\rm M_\odot}$ have the three axes with the same length within 10\,\% or less \citep[e.g.,][]{2014MNRAS.443.3208D,2019Galax...7...81Z}. Thus, deviations from spherical symmetry are neglected when measuring the total density $\rho$ consistent with the observed circular velocity (Sect.~\ref{sec:observations}, Eq.~[\ref{eq:vc2rho}]).  However, for the sake of comprehensiveness and because it is simple to treat, we next discuss the effect of having the total effective Newtonian mass distributed in a thin disk. Under this assumption and close to the center of the potential, we approximate the total surface density  as 
\begin{equation}
  \Sigma(R)  \propto R^{-\delta},
  \label{eq:sigmadelta}
\end{equation}
so that in the MOND regime where $\gobs\propto \sqrt{\gbar}$ (Sect.~\ref{sec:intro}),  the indexes for the baryon surface density and total mass are related as
\begin{equation}
  \delta = \beta/2.
  \label{eq:betadelta}
\end{equation}
On the other hand, the equivalence between true surface density and an effective volume density in Eq.~(\ref{eq:transformation}) implies
$\rho\simeq r^{-(1+\delta)}$ so that 
\begin{equation}
\frac{d\log\rho}{d\log r}\simeq -(1+\delta).
\end{equation}
Therefore,   for the total volume density profile to have a core (Eq.~[\ref{eq:coredef}]), $1+\delta \simeq 0$.  Thus, given the constraint set by Eq.~(\ref{eq:betadelta}), a core requires $\beta\simeq -2$ and, consequently, a pronounced drop of  $\sigmabardisk$ when $R\rightarrow 0$   (Eq.~[\ref{eq:sigmadisk}]).

 \subsection{Surface density corresponding to volume densities with central drops}\label{sec:rho2sigma}

\begin{figure}
  \centering
\includegraphics[scale=0.55]{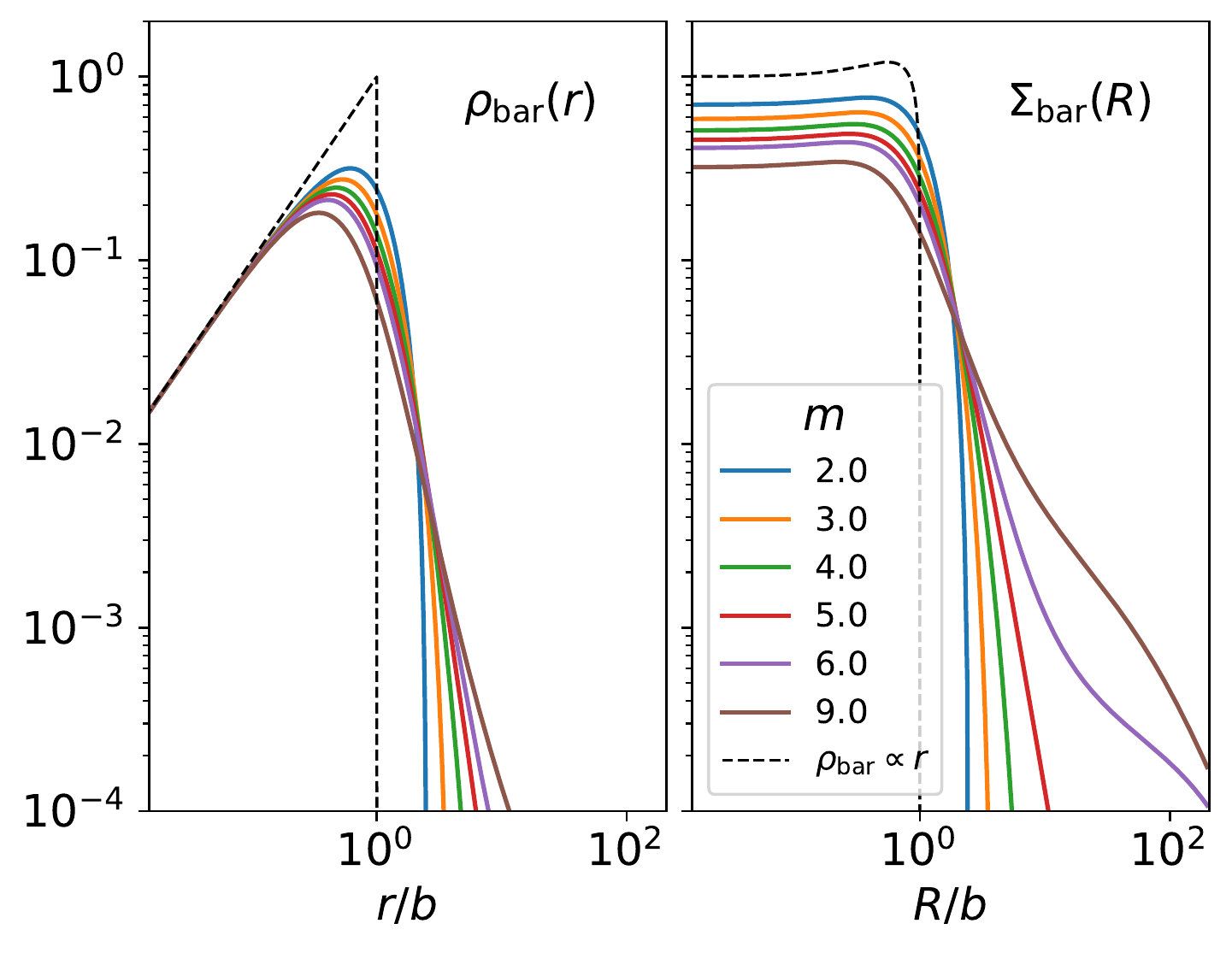}
\caption{Left panel: hollow polytropes (Eq.~[\ref{eq:hollowpoly}]) that produce a total density profile with a core. Right panel:  Abel transform of the 3D profiles shown in the left panel. The color code of both panels is the same and gives the polytropic index $m$. The figure also includes the limiting case of a pure linearly growing volume density profile and its corresponding surface density (the dashed lines, given by Eq.~[\ref{eq:this}] with $\rho_0=1$ and $r_0=b$). Densities are given in arbitrary units.
}
\label{fig:lane_emden_grid3_plot}
\end{figure}

 Some of the observations discussed below have to do with surface density rather than volume density. The question arises as to what is the baryon surface density corresponding to the condition of cores following RAR (Eq.~[\ref{eq:power}]). The response does not have a closed analytic expression since the surface density depends not only on the central density but also on the full density profile through its projection in a plane (i.e., through its Abel transform). Thus, the corresponding surface density has to be computed numerically on a case-by-case basis. Figure~\ref{fig:lane_emden_grid3_plot} shows $\sigmabar(R)$ for a number hollow polytropes. They were computed numerically through the Abel transform of their 3D distribution given in Eq.~(\ref{eq:hollowpoly}).  $\sigmabar(R)$ is shown in the right panel whereas the  $\rhobar(r)$ from which they derive are displayed in the left panel. The resulting $\sigmabar$  profiles have an small inner depression with an upturn at $R=b$.
The plateau with subsequent upturn is a generic prediction of volume densities with a central drop, as can be drawn from the following argument. Adopting the density in Eq.~(\ref{eq:power}) for $r\leq r_0$ and 0 elsewhere, $\sigmabar(R)$ turns out to be analytic. A simple integration yields
\begin{equation}
  \sigmabar(R) = \frac{r_0\rho_0}{X^2}\Big[X\sqrt{X^2-1}+\ln\Big(X+\sqrt{X^2-1}\Big)\Big],
  \label{eq:this}
\end{equation}
with $X=r_0/R$. Equation~(\ref{eq:this}) holds for $R\leq r_0$ while $\sigmabar(R)=0$  elsewhere. Figure~\ref{fig:lane_emden_grid3_plot} includes the surface density in this limiting  case  (the black dashed line in the right panel), which displays the same plateau plus upturn of the hollow polytropes (the color lines). Thus, any other profile with a core approximately given by Eq.~(\ref{eq:power}) is expected to share this feature.

%
\section{Observational constraints}\label{sec:observations}

A cornerstone of our argumentation is the fact that dwarf galaxies often have a inner core. Their total mass profile shows a central plateau when the RCs are interpreted in terms of DM \citep[e.g.,][]{1990AJ....100..648J,1995ApJ...447L..25B,2014ApJ...789...63A,2015AJ....149..180O,2020ApJ...902...98G}. The observed density profile is obtained keeping in mind that, for spherically symmetric system, the observed circular velocity $V_c$ is just a mapping of $\gobs$ \citep[e.g.,][]{2008gady.book.....B}, specifically,
\begin{equation}
  \gobs = \frac{V_c^2}{r},
  \label{eq:vc}
\end{equation}
which leads to \citep[e.g.,][]{2001ApJ...552L..23D},
\begin{equation}
  \rho(r) = \frac{1}{4\pi G}\,\frac{V_c^2}{r^2}\,\Big[1+2\frac{d\log V_c}{d\log r}\Big].
  \label{eq:vc2rho}
\end{equation}
Examples of such cores are given in Fig.~\ref{fig:little_things2b}, top panel.  It displays the 26 Little Things (LT) galaxies \citep{2012AJ....144..134H,2012AJ....143...47Z}, which are nearby ($\gtrsim 10.3${Mpc) low-mass dwarfs ($M_\star \sim 10^{7-8} \,{\rm M}_\odot$; see Table~\ref{tab:appa}) with well resolved H{\sc i} RCs \citep{2015AJ....149..180O}. Other excellent data sets used to study RAR \citep[e.g., SPARC;][]{2018A&A...615A...3L} have galaxies reaching the Mondian regime only in the outskirts and so are not suitable for our study. The LT dataset includes gas mass and stellar mass profiles, and so, all the ingredients to compute $\gobs$ and $\gbar$ independently.  Figure~\ref{fig:little_things2b} shows $\rho$ (top panel) and $V_c$ (bottom panel) for these galaxies.  As usual, the profiles have been normalized to the radius $r_{0.3}$ where the logarithmic derivative of the RC is 0.3, namely,
\begin{equation}
  \frac{d\log V_c}{d\log r}\Big(r_{0.3}\Big)= 0.3.
  \label{eq:03}
\end{equation}
Note that, overall, the densities tend to a constant value for $r\le r_{0.3}$, i.e., they show the core characteristics of the LT galaxies \citep{2015AJ....149..180O,2020A&A...642L..14S}. For reference,  Fig.~\ref{fig:little_things2b} includes a NFW profile and its RC (the black dashed line), as well as a polytrope of order 5 and its RC (the blue solid line).  The polytrope reproduces very well both $\rho$ and $V_c$, despite there is no degree of freedom in the comparison  \citep[for a discussion on this, see][]{2020A&A...642L..14S}.
\begin{figure*}
  \centering
\includegraphics[scale=0.8]{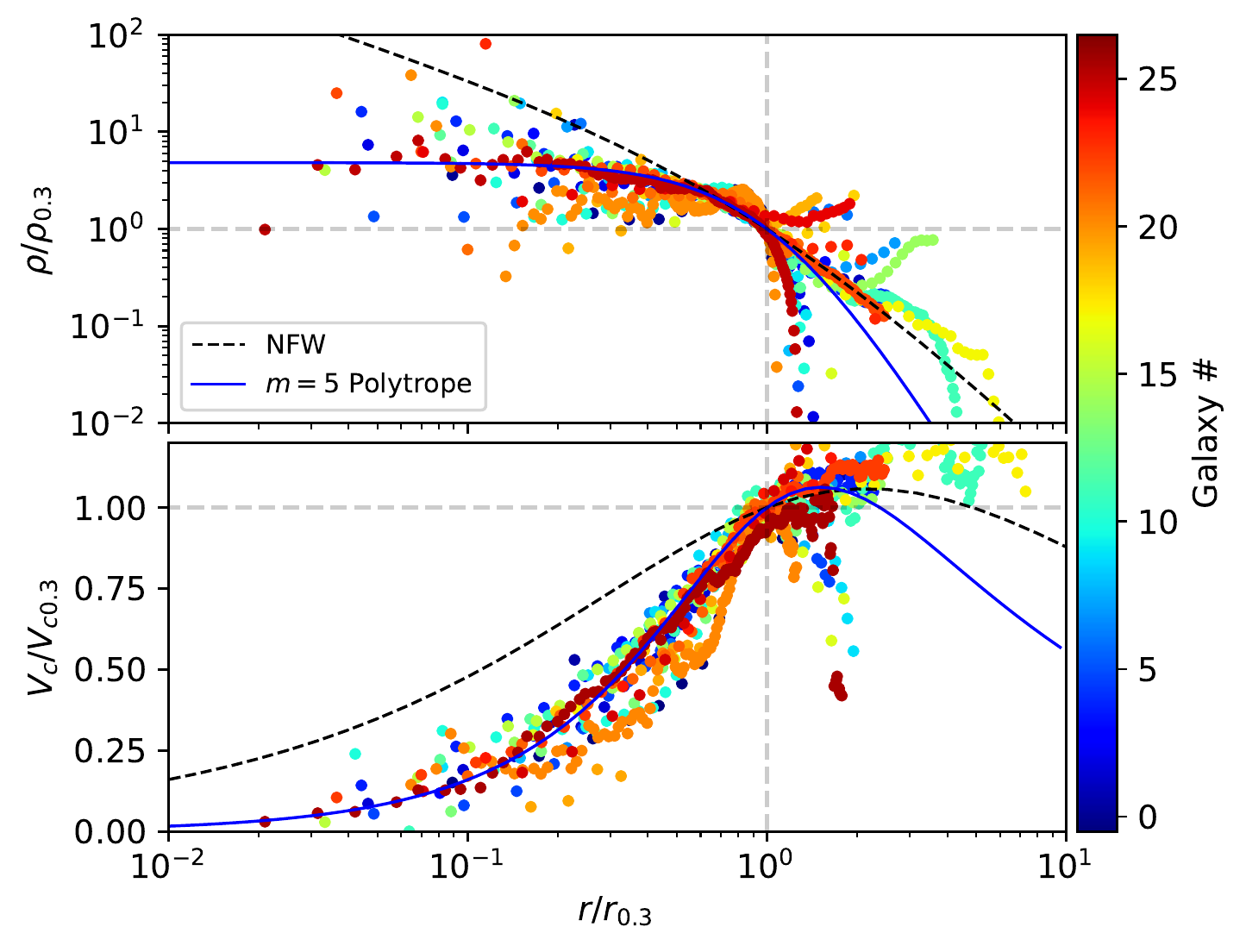}
\caption{LT galaxies from \citet{2015AJ....149..180O}. Top panel: total density $\rho$ inferred from the circular velocity $V_c$ through Eq.~(\ref{eq:vc2rho}). Bottom panel: $V_c$ inferred from the observed RCs  corrected by asymmetric drift. The data are normalized to $r_{0.3}$, where the logarithmic derivative of the RC equals 0.3 (Eq.~[\ref{eq:03}]). Each color represents a different galaxy with the identifier given in the color bar (see Table~\ref{tab:appa} for the equivalence with galaxy names). For reference, the plot includes a NFW profile and the corresponding RC (the black dashed lines), as well as the density and velocity of a polytrope of order $m=5$ (the blue solid lines).  
}
\label{fig:little_things2b}
\end{figure*}
%
%

What happens with the distribution of baryons in these dwarfs with cores? 
Low-mass dwarfs are far from being represented by thin disks. The baryons are distributed over the full 3D volume rather than concentrated in a plane. Their stellar mass distribution tends to be triaxial, with the axial ratio between the minor and the major axes typically reaching 0.5 \citep[e.g.,][]{2019MNRAS.486L...1S,2019ApJ...883...10P}. Moreover, the random velocities of the stars and gas are usually comparable to the rotational velocities \citep[e.g.,][]{2013ApJ...767...74S,2017ApJ...834..181O} implying that the motions cannot be confined within a thin disk. This large random velocity is also characteristic of the H{\sc i} distribution of the LT galaxies, for which random motions have been measured to be similar to rotation and other noncircular motions \citep[][]{2012AJ....144..134H,2019AJ....158...23H}. Thus, the spherical symmetry assumed in Sect.~\ref{sec:maineqs} seems to be a reasonable approximation for dwarfs with cores, keeping in mind that the alternative geometry of being thin disks produces central density drops even more pronounced (Sect.~\ref{sec:diskbar}). On the other hand, 
the observed mass {\em surface} density characteristic of dwarf galaxies tends to show a central plateau. This happens with the stellar mass distribution \citep[e.g.,][]{2021ApJ...921..125S,2021ApJ...922..267C}, but also with the gas mass \citep[e.g.,][]{2021AJ....161...71H}. However, the observed surface density profiles are quite different from the ones resulting from projecting in the plane of the sky  3D density profiles with central hollows  (Sect.~\ref{sec:rho2sigma}). The observed inner surface density profile does not show the flat-top plus upturn shape to be expected in this case (Fig.~\ref{fig:lane_emden_grid3_plot}, right panel).
Thus, the existing observations do not suggest the low-mass dwarfs to have a central drop in their baryon density. 

Since the typical dwarfs with cores do not have obvious drops in baryon density, it is to be expected that their central regions often elude RAR.
This conjecture can be checked out directly with the LT galaxies.  Using the circular velocities kindly provided by S.-H.~Oh, and through the help of Eq.~(\ref{eq:vc}), one can plot the LT galaxies in the RAR plane. This is shown in Fig.~\ref{fig:little_things2e}, where the $V_c$ rendering $\gobs$ has been corrected for asymmetric drift \citep{2015AJ....149..180O}, and the baryon RC includes the contribution of both gas and stars.\footnote{We also repeat the exercise using gas mass alone, which is the dominant baryon component for most LT galaxies at all radii. The results are very much in agreement with those shown and discussed in the main text.} Note that $\gobs$ and $\gbar$ are independent of $r_{0.3}$, and depend only on the observed RC and the mass distribution inferred from H{\sc i} maps (gas) and from near IR images (stars). As it is clear from  Fig.~\ref{fig:little_things2e}, top panel, many LT galaxies do deviate from RAR. This deviation is larger in the innermost regions with  $r\lesssim r_{0.3}$ (Fig.~\ref{fig:little_things2e}, bottom panel).
\begin{figure*}
  \centering
\includegraphics[scale=0.8]{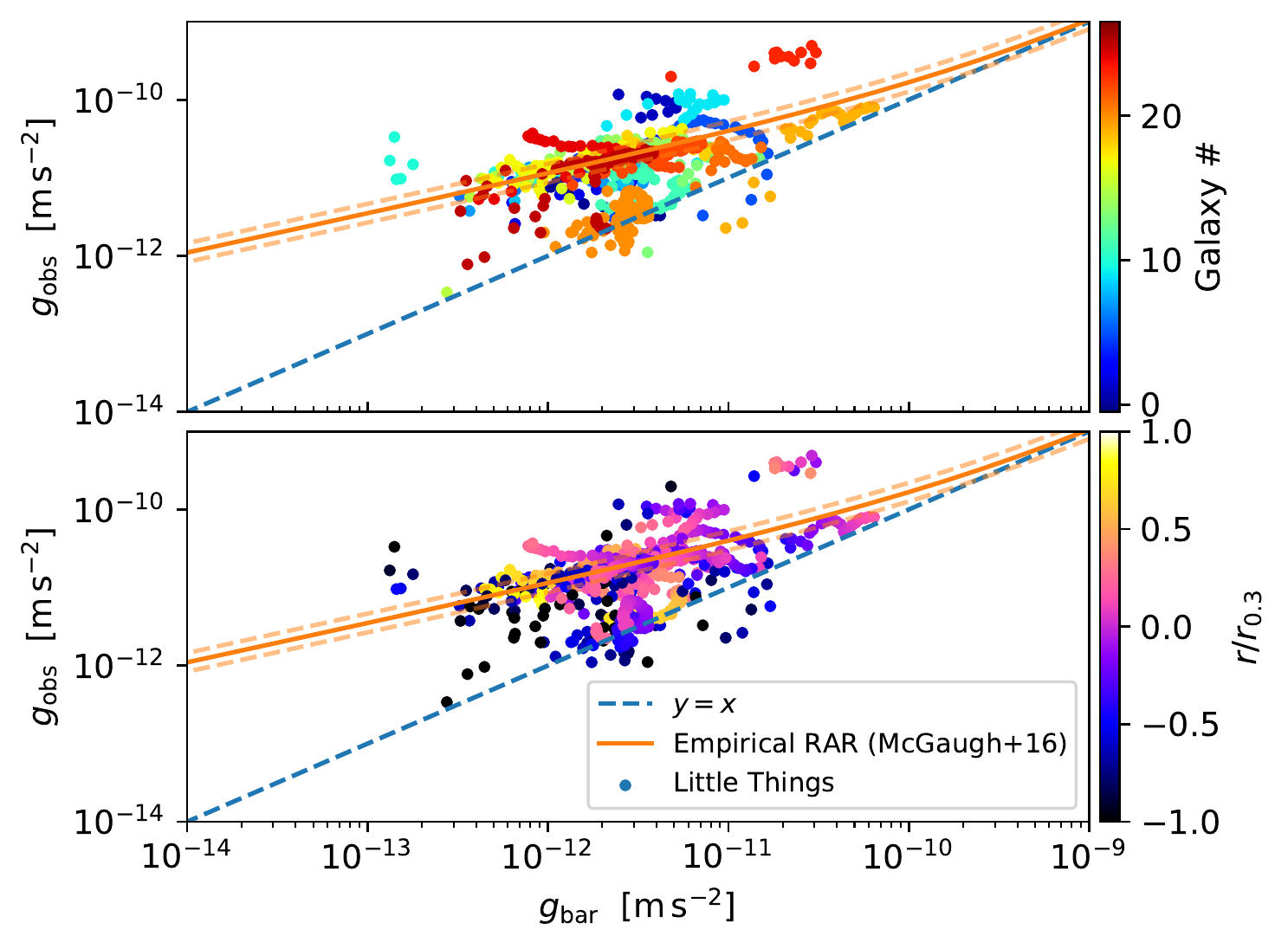}
\caption{LT galaxies in the RAR plane. Top panel: all 26 LT galaxies color coded with galaxy \# (and arbitrary identifier; see Table~\ref{tab:appa}). Bottom panel: same plot but color coded with the normalized distance to the galaxy center (the same normalization to $r_{0.3}$ used for Fig.~\ref{fig:little_things2b}). The solid orange lines represent the RAR as given by Eq.~(\ref{eq:rar})
}
\label{fig:little_things2e}
\end{figure*}
To be more precise, we separate the 26 LT galaxies into those that do not follow RAR (13 objects: Fig.~\ref{fig:little_things2ff}, top panel), and those that may follow RAR (another 13 objects: Fig.~\ref{fig:little_things2ff}, bottom panel; see also Table~\ref{tab:appa}). Figure~\ref{fig:little_things2ff} includes the error bars propagated from the observational error in the RCs.  Note the large deviations from the RAR in  Fig.~\ref{fig:little_things2ff}, top panel, which are systematic within individual galaxies. The galaxies denoted as not following RAR clearly deviate from the theoretical relation even when the error bars are considered. 
\begin{figure*}
  \centering
\includegraphics[scale=0.8]{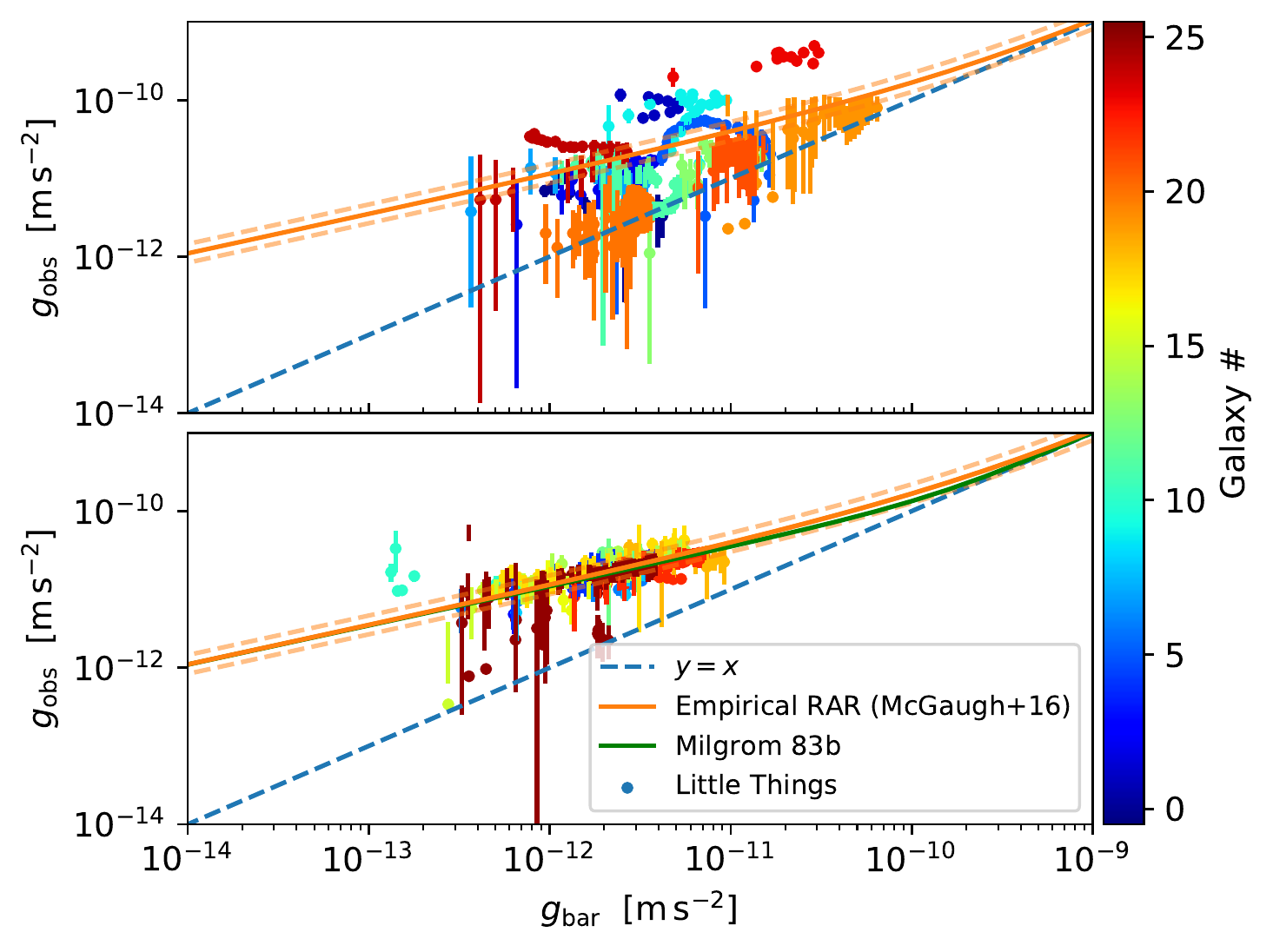}
\caption{LT galaxies in the RAR plane. Top panel: LT galaxies which do not follow RAR, color coded with galaxy \# as in the top panel of Fig.~\ref{fig:little_things2e} (see also Table~\ref{tab:appa}). Bottom panel: LT galaxies approximately following RAR, also color coded with galaxy number. The plots include error bars propagated from the observational error in the RCs. The barely visible green solid line illustrates the kind of differences existing between different parameterizations of RAR (this one is given in Eq.~[\ref{eq:appb3}]).
} 
\label{fig:little_things2ff}
\end{figure*}

We have not been able to find anything suspicious or even special in the RC of the LT galaxies that deviate from RAR. 
Figure~\ref{fig:little_things2g} shows the RCs of the LT galaxies split into those not following RAR (top panels) and those following RAR (bottom panels).  They are also separated into total RCs (left panels) and RCs for baryons only (right panels). No systematic differences are noticeable.  
\begin{figure*}
\includegraphics[scale=0.6]{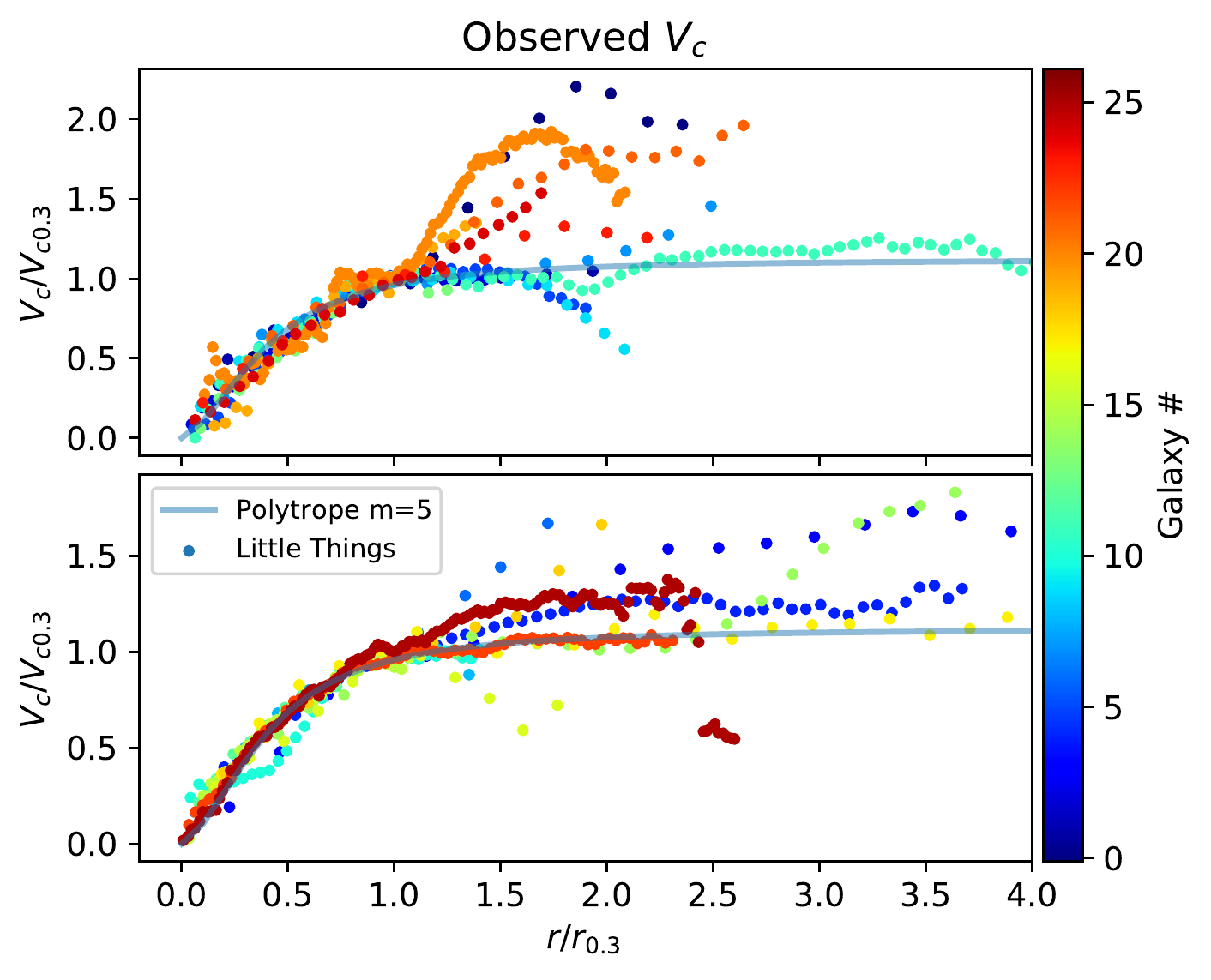}
\includegraphics[scale=0.6]{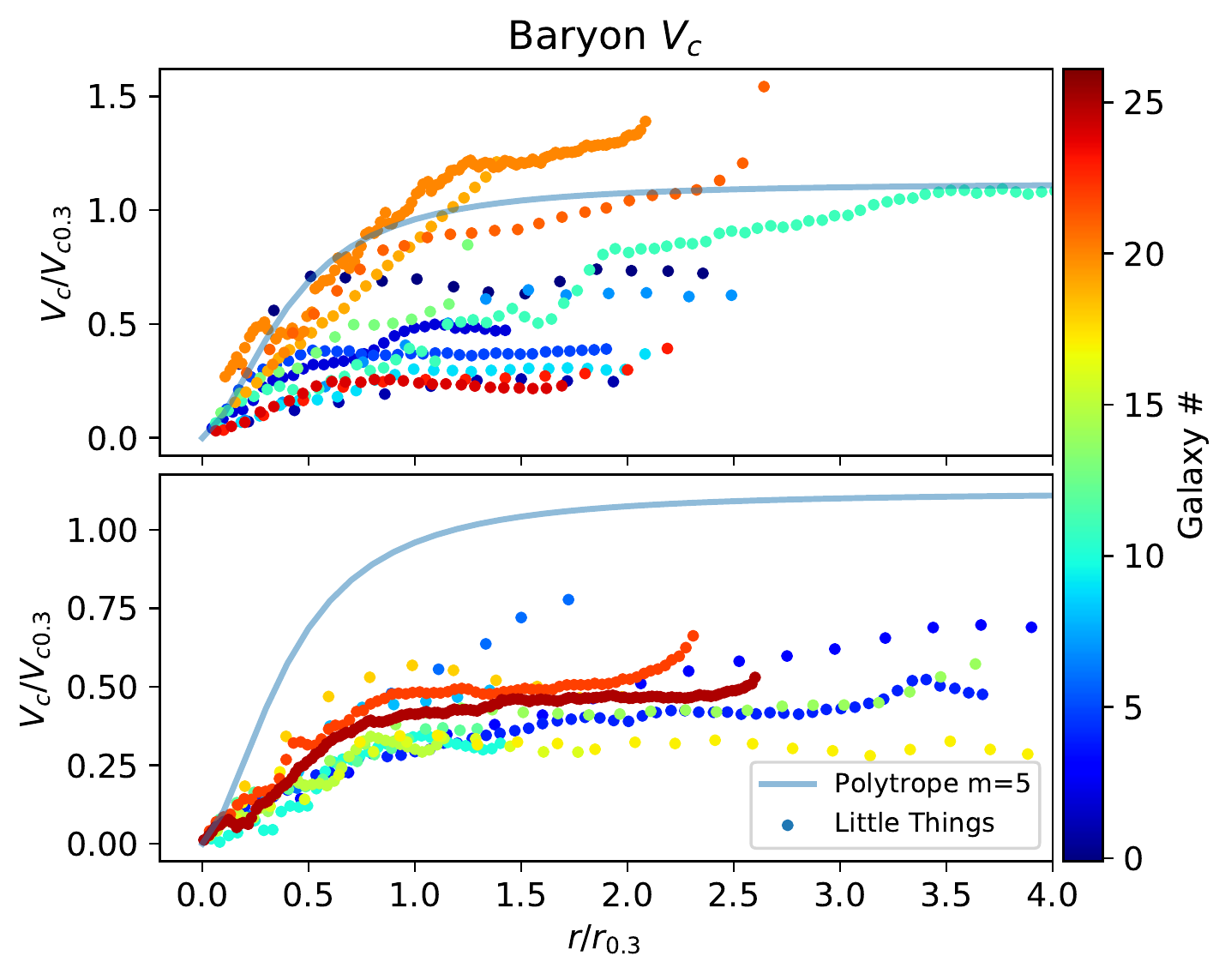}
\caption{LT galaxies total observed RCs (left panels) and baryon-only RCs (right panels). They were used to compute $\gobs$ and $\gbar$ in Fig.~\ref{fig:little_things2ff}. The panels on top show  the  LT galaxies which do not follow RAR, whereas the bottom panels include LT galaxies following RAR. They are color coded with galaxy \# as indicated in the bar and as given Figs.~\ref{fig:little_things2e} and \ref{fig:little_things2ff}, and in Table~\ref{tab:appa}. 
}
\label{fig:little_things2g}
\end{figure*}

\section{Discussion and conclusions}\label{sec:conclusions}

In dwarf  galaxies, the gravity is often within the Mondian regime at all radii (Sect.~\ref{sec:intro}). Some of these dwarf galaxies also show cores, in the sense that assuming Newton's gravity to explain their RCs, the resulting total density profile presents a central plateau or core ($d\log \rho/d\log r\rightarrow 0$ when $r\rightarrow 0$). If MOND rules gravity, the existence of this core implies a baryon content whose density decreases toward the center of the gravitational potential ($\rhobar\rightarrow 0$ when $r\rightarrow 0$; Sect.~\ref{sec:maineqs}).
We derive this result assuming the MOND gravitational law to be given by the RAR expression in Eq.~(\ref{eq:rar}), however, the same result still holds for any other MOND parameterization since all of them agree in the Mondian regime (Sect.~\ref{sec:maineqs}).

Observations do not favor the presence of a drop in baryon density at the center of dwarf galaxies  (Sect.~\ref{sec:observations}). Baryon radial density profiles, whether they are tracing gas or stars, tend to show central  plateaus rather than central drops. Moreover, galaxies with well-measured H{\sc i}~RCs and baryon profiles often deviate from RAR (Fig.~\ref{fig:little_things2ff}, top panel).  The mismatch remains even when observational errors are taken into account.

One might think of various workarounds to explain why the predicted central baryon drop  is not observed even if MOND holds.  One such possibility is the breakdown of the spherical symmetry assumed in our discussion (Sect.~\ref{sec:maineqs}). However, this solution is not expected to work because when all baryons are concentrated in a thin disk, which completely breaks down the spherical symmetry, the central baryon drop becomes even more necessary ($\rhobar$ must be proportional to $r^2$ rather than $r$; Sect.~\ref{sec:diskbar}). Although the general case of a triaxial baryon distribution cannot be treated analytically, it is anticipated to remain in between the two extremes (i.e., spheroids and disks) and so to retain the need for a density drop.

Another alternative solution could be that all dwarfs deviating from RAR are not in mechanical equilibrium. The analyzed observations would have captured them during a transient, when motions are not set by the force balance expected from MOND. This explanation cannot be fully discarded but it seems unlikely due to the large frequency of exceptions to MOND  (50\% in the galaxy sample studied in Sect.~\ref{sec:observations}) and the short time-scale required to reach mechanical balance \citep[of the order of a few hundred Myr; see, e.g., ][]{2021MNRAS.505.4655S}.

A third way to overcome the problem could be the presence of a systematic bias in the measured RCs \citep[the claim that cores could often result from artifacts, e.g.,][]{2022arXiv220316652R}. Once again, this is inconsistent with the observed properties of the LT family. Similar galaxies have been measured with the same instrumentation and under similar conditions and, nevertheless,  some members show cores while others do not (cf. top and bottom panels in Fig.~\ref{fig:little_things2ff}).  

An independent argument reinforces the difficulty of having MOND self-gravitating structures that lack baryons at their centers.  MOND gravity is also a central force, thus, it should not produce spherically symmetric structures where the source of the force (i.e., the baryons) are lacking toward the center of the gravitational potential. Even if they could be formed, the resulting structure should be Rayleigh-Taylor-like unstable. This difficulty is confirmed by the few existing numerical simulations of galaxy formation and evolution using MOND acceleration. The central density drop is missing in all simulated galaxies, whether they are massive \citep{2014A&A...571A..82C,2018A&A...614A..59B}, intermediate-mass \citep{2021MNRAS.503.2833R}, or even dwarfs \citep[][$M_\star\simeq 2\times 10^8\,{\rm M}_\odot$]{2021A&A...653A.170B}. Baryons are present in the centers of spheroids as well as in disks \citep{2007A&A...464..517T,2020ApJ...890..173W}.

We have shown that RAR is often disobeyed at the center of dwarf galaxies. This inconsistency between RAR and observations has a  totally different  consequence for cold DM and for MOND. For cold DM, RAR is just a secondary relation emerging from the interplay between baryons and DM \citep[e.g.,][]{2017MNRAS.471.1841N,2017PhRvL.118p1103L}. For MOND, however, RAR is a law of nature that always holds.
    Thus, the exceptions to RAR fit easily within the cold DM paradigm as particular cases where the secondary relationship did not arise, yet pose a more fundamental problem for MOND.

\begin{acknowledgments}

Thanks are due to Se-Heon~Oh for providing the RCs of the LT galaxies in computer readable format.
Dedrie Hunter and Bruce Elmegreem were extremely helpful providing insight on the LT RCs.
Thanks are also due to Andr\'es~Castillo,  Jorge~Mart\'\i n Camalich, and Andr\'es Ba\~nares for discussions on the true shape of the LT RCs, and to Claudio Dalla Vecchia for a careful reading of the manuscript.
The comments of an anonymous referee helped us to sharpen some of the arguments.
This research was partly funded by the Spanish Ministry of Science and Innovation, project  PID2019-107408GB-C43 (ESTALLIDOS), and by Gobierno de Canarias through EU FEDER funding, project PID2020010050.

\end{acknowledgments}




\software{astropy \citep{2013A&A...558A..33A,2018AJ....156..123A}} 

\appendix

\section{Table with galaxy number identification}\label{app:appa}

This appendix provides the correspondence between the galaxy numbers using for plotting (Figs.~\ref{fig:little_things2b}, \ref{fig:little_things2e}, \ref{fig:little_things2ff}, and \ref{fig:little_things2g}) and the LT galaxy name as given by \citet{2015AJ....149..180O} in their Table~2.  

\begin{table}
	\centering
	\caption{Galaxy number identification.}
	\label{tab:appa}
	\begin{tabular}{ccccc} 
		\hline
          Galaxy \# & Name & RAR? & $M_\star$&$M_g$ \\
          (1)&(2)&(3)&(4)&(5)\\
          \hline
0 & CVnIdwA & N & 0.49 & 2.91 \\
1 & DDO101 & N & 6.54 & 3.48 \\
2 & DDO126 & N & 1.62 & 16.4 \\
3 & DDO133 & Y & 3.04 & 12.9 \\
4 & DDO154 & Y & 0.83 & 35.3 \\
5 & DDO168 & N & 5.85 & 25.9 \\
6 & DDO210 & Y & 0.06 & 0.14 \\
7 & DDO216 & N & 1.51 & 0.49 \\
8 & DDO43 & Y & \dots & 23.3 \\
9 & DDO46 & N & \dots & 22.1 \\
10 & DDO47 & Y & \dots & 46.8 \\
11 & DDO50 & N & 10.6 & 132. \\
12 & DDO52 & Y & 5.31 & 33.4 \\
13 & DDO53 & N & 0.97 & 7.00 \\
14 & DDO70 & Y & 1.96 & 3.80 \\
15 & DDO87 & Y & 3.27 & 29.1 \\
16 & F564-V3 & Y & \dots & 4.37 \\
17 & Haro29 & Y & 1.43 & 9.35 \\
18 & Haro36 & Y & \dots & 11.2 \\
19 & IC10 & N & \dots & 1.65 \\
20 & IC1613 & N & 2.88 & 5.93 \\
21 & NGC1569 & N & 36.9 & 20.2 \\
22 & NGC2366 & Y & 6.94 & 108. \\
23 & NGC3738 & N & 12.6 & 46.6 \\
24 & UGC8508 & N & 0.77 & 1.19 \\
25 & LWM & Y & 1.62 & 11.2 \\
		\hline
	\end{tabular}
{\small
        \begin{tabular}{l} 
	  (1) LT galaxy \# used for plotting.\\
          (2) Galaxy name as given by \citet{2015AJ....149..180O} in their Table~2.\\
          (3) Does it follow RAR?\\
          (4) Stellar mass in units of $10^7\,{\rm M}_\odot$ \citep[][Table~2]{2015AJ....149..180O}.\\
          (5) Gas mass in units of $10^7\,{\rm M}_\odot$ \citep[][Table~2]{2015AJ....149..180O}.\\
	\end{tabular}
}
\end{table}





\end{document}